\documentclass[runningheads]{llncs}
\usepackage{graphicx}
\usepackage{multirow}
\usepackage[T1]{fontenc}
\usepackage{graphicx}
\begin{document}
\title{Characterization of pedestrian contact interaction trajectories}
%
%
\author{
Jaeyoung Kwak\inst{1}\and
Michael H. Lees\inst{2} \and
Wentong Cai\inst{1}
}
\authorrunning{J. Kwak \textit{et al.}}
%
\institute{Nanyang Technological University, Singapore 639798, Singapore\\
\email{\{jaeyoung.kwak,aswtcai\}@ntu.edu.sg}\\
\and
University of Amsterdam, Amsterdam 1098XH, The Netherlands\\
\email{m.h.lees@uva.nl}}
\maketitle              
\begin{abstract}
A spreading process can be observed when a particular behavior, substance, or disease spreads through a population over time in social and biological systems. It is widely believed that contact interactions among individual entities play an essential role in the spreading process. Although the contact interactions are often influenced by geometrical conditions, little attention has been paid to understand their effects especially on contact duration among pedestrians. To examine how the pedestrian flow setups affect contact duration distribution, we have analyzed trajectories of pedestrians in contact interactions collected from pedestrian flow experiments of uni-, bi- and multi-directional setups. Based on standardized maximal distance, we have classified types of motions observed in the contact interactions. We have found that almost all motion in the unidirectional flow setup can be characterized as subdiffusive motion, suggesting that the empirically measured contact duration tends to be longer than one estimated by ballistic motion assumption. However, Brownian motion is more frequently observed from other flow setups, indicating that the contact duration estimated by ballistic motion assumption shows good agreement with the empirically measured one. Furthermore, when the difference in relative speed distributions between the experimental data and ballistic motion assumption is larger, more subdiffusive motions are observed. This study also has practical implications. For instance, it highlights that geometrical conditions yielding smaller difference in the relative speed distributions are preferred when diseases can be transmitted through face-to-face interactions. 

\keywords{Pedestrian flow \and Contact interaction \and Brownian motion \and Subdiffusive motion \and Contact duration.}
\end{abstract}
\section{Introduction}
\label{section:intro}

Modeling contact interactions among individual entities is essential to understand spreading processes in social and biological systems, such as information diffusion in human populations~\cite{Samar_IEEE2006,Wu_IEEE2008} and transmission of infectious disease in animal and human groups~\cite{Hu_2013,Manlove_2022}. For the spreading processes in social and biological systems, one can observe a contact interaction when two individual entities are within a close distance, so they can exchange substance and information or transmit disease from one to the other one. In previous studies, macroscopic patterns of contact interactions are often estimated based on simple random walking behaviors including ballistic motion. For example, Rast~\cite{Rast_PRE2022} simulated continuous-space-time random walks based on ballistic motion of non-interacting random walkers. Although such random walk models have widely applied to estimate contact duration for human contact networks, little work has been done to study the influence of pedestrian flow geometrical conditions on the distribution of contact duration.

To examine how the geometrical conditions of pedestrian flow affect the contact duration distribution, we perform trajectory analysis for the experimental dataset collected from a series of experiments performed for various pedestrian flow setups. The trajectory analysis of moving organisms, including proteins in living cells, animals in nature, and humans, has been a popular research topic in various fields such as biophysics~\cite{Saxton_1997,Manzo_2015,Shen_2017}, movement ecology~\cite{Benhamou_2007,Edelhoff_2016,Getz_PNAS2008}, and epidemiology~\cite{Rutten_SciRep2022,Wilber_2022}. Single particle tracking (SPT) analysis, a popular trajectory analysis approach frequently applied in biophysics and its neighboring disciplines, characterizes the movement dynamics of individual entities based on observed trajectories~\cite{Manzo_2015,Qian_1991}. According to SPT analysis, one can identify different types of diffusion, for instance, directed diffusion in which individuals move in a clear path and confined diffusion in which individuals tend to move around the initial position. The most common method for identifying diffusion types is based on the mean-squared displacement (MSD), which reflects the deviation of an individual's position with respect to the initial position after time lag~\cite{Qian_1991,Michalet_PRE2010}. Motion types can be identified based on the diffusion exponent. MSD has been widely applied for various trajectory analysis studies in biophysics~\cite{Goulian_2000,Hubicka_PRE2020}. For pedestrian flow trajectory analysis, Murakami~\textit{et~al.}~\cite{Murakami_interface2019,Murakami_SciAdv2021} analyzed experimental data of bidirectional pedestrian flow and reported diffusive motion in individual movements perpendicular to the flow direction. They suggested that uncertainty in predicting neighbors' future motion contributes to the appearance of diffusive motion in pedestrian flow. 

Previous studies have demonstrated usefulness of SPT analysis in examining movement of individuals. However, SPT analysis does not explicitly consider relative motions among individuals in contact, suggesting that analyzing the relative motions can reveal patterns that might not be noticeable from the SPT analysis approach. For example, if two nearby individuals are walking in parallel directions together with a similar speed, one might be able to see various shapes of relative motion trajectories although the individual trajectories are nearly straight lines. For contact interaction analysis, the analysis of relative motion trajectories can be utilized to predict the length of contact duration and identify contact interaction characteristics such as when the interacting individuals change walking direction significantly. Regarding the spreading processes, understanding of relative motion trajectory can be applied to identify optimal geometrical conditions that can minimize contract duration when diseases can be transmitted through face-to-face interactions. 

Although MSD is simple to apply, MSD has limitations. Refs.~\cite{Briane_2016,Briane_PRE2018,Janczura_PRE2020} pointed out that MSD might not be suitable for short trajectories to extract meaningful information. Additionally, due to its power-law form, the estimation of the diffusion exponent in MSD is prone to estimation errors~\cite{Kepten_PLOS2015,Burnecki_SciRep2015}. As an alternative to MSD, various approaches have been proposed, including the statistical test approach~\cite{Briane_2016,Briane_PRE2018,Weron_PRE2019,Janczura_2022} and machine learning approach~\cite{Kowalek_JPhysA2022}.

In this work, we analyze trajectories collected from different experiment setups including uni-, bi-, and multi-directional flow for pedestrian contact interactions. Rather than using individual trajectories, we analyze the relative motion of interactions to understand {\em pedestrian contact interactions}. We identify different types of motion observed in pedestrian contact interactions based on a statistical test procedure. For the statistical test procedure, we measured a standardized value of largest distance traveled by an individual from their starting point during the contact interaction. Our results demonstrate that examining the interactions in this way can provide important insights regarding contact duration, and hence help estimate transmission risk in different pedestrian flow conditions. 

The remainder of this paper is organized as follows. Section~\ref{section:dataset} describes the datasets including pedestrian flow experiment setups and some descriptive statistics. The statistical test procedure and trajectory classification results are presented in Section~\ref{section:analysis}. We discuss the findings of our analysis in Section~\ref{section:conclusion}.

\section{Datasets}
\label{section:dataset}

\begin{figure}
	\centering
	\includegraphics[width=1.05\textwidth]{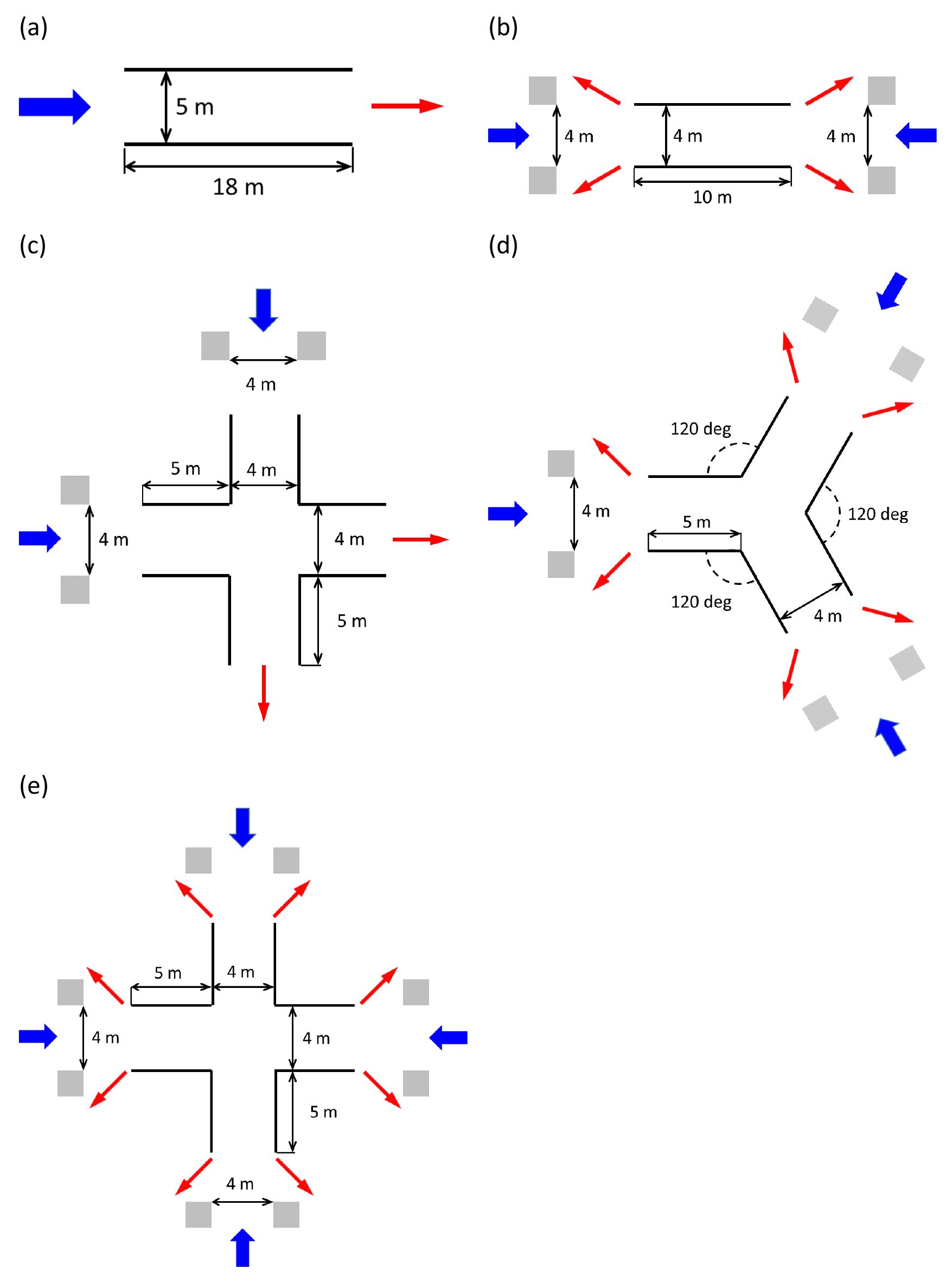}
	\vspace*{-5mm}
	\caption{Schematic representation of experiment setups: (a) unidirectional flow (scenario name: uni-05), (b) bidirectional flow (scenario name: bi-b01), (c) 2-way crossing flow (scenario name: crossing-90-d08), (d) 3-way crossing flow (scenario name: crossing-120-b01), and (e) 4-way crossing flow crossing flow (scenario name: crossing-90-a10). Here, blue thick arrows show the walking direction of incoming pedestrians entering the corridors and red thin arrows indicates the walking direction of outgoing pedestrians leaving the corridors. A pair of gray rectangles is placed to set up an entrance of pedestrian group entering the corridor.} 
	\label{fig:ExperimentSetup}
\end{figure}

Figure~\ref{fig:ExperimentSetup} shows the sketches of various experiment setup: uni-directional flow, bi-directional flow, 2-way crossing flow, 3-way crossing flow, and 4-way crossing flow. In the uni-directional flow setup, pedestrians were walking to the right in a straight corridor of 5~m wide and 18~m long. In a bi-directional flow, two groups of pedestrians were entering a straight corridor of 4~m wide and 10~m long through 4~m wide entrance and then walking opposite directions. They left the corridor through the open passage once they reached the other side of the corridor. In 2-way, 3-way, and 4-way crossing flows, different groups of pedestrians were entering the corridor through 4~m wide entrance and walked 5~m before and after passing through an intersection (4~m by 4~m rectangle in 2-way crossing and 4-way crossing flows, and 4~m wide equilateral triangle in 3-way crossing flow). Similar to the setup of bi-directional flow, pedestrian groups left the corridor through the open passage after they reached the end of corridors. A more detailed description of the experiment setups can be found in Refs.~\cite{Holl_Dissertation2016,Cao_JSTAT2017,url_dataset}.

\begin{table}
	\caption{Basic descriptive statistics of representative scenarios.}
	\label{table:BasicStatistics}
	\setlength{\tabcolsep}{8pt}
	\centering
	\resizebox{11.5cm}{!}{
		\begin{tabular}{c*{5}{r}}
			\hline
			Setup & Scenario name & N & Period (s) & No. contacts\\
			\hline
			Uni-directional & uni-05 & 905 & 157.68 & 25390\\
			Bi-directional & bi-b10 & 736 & 324.56 & 57126\\
			2-way crossing & crossing-90-d08 & 592 & 147.88 & 41648\\
			3-way crossing & crossing-120-b01 & 769 & 215.63 & 93156\\
			4-way crossing & crossing-90-a10 & 324 & 94.08 & 9270\\
			\hline
		\end{tabular}
	}
\end{table}

\begin{figure}
	\centering
	\includegraphics[width=1.0\textwidth]{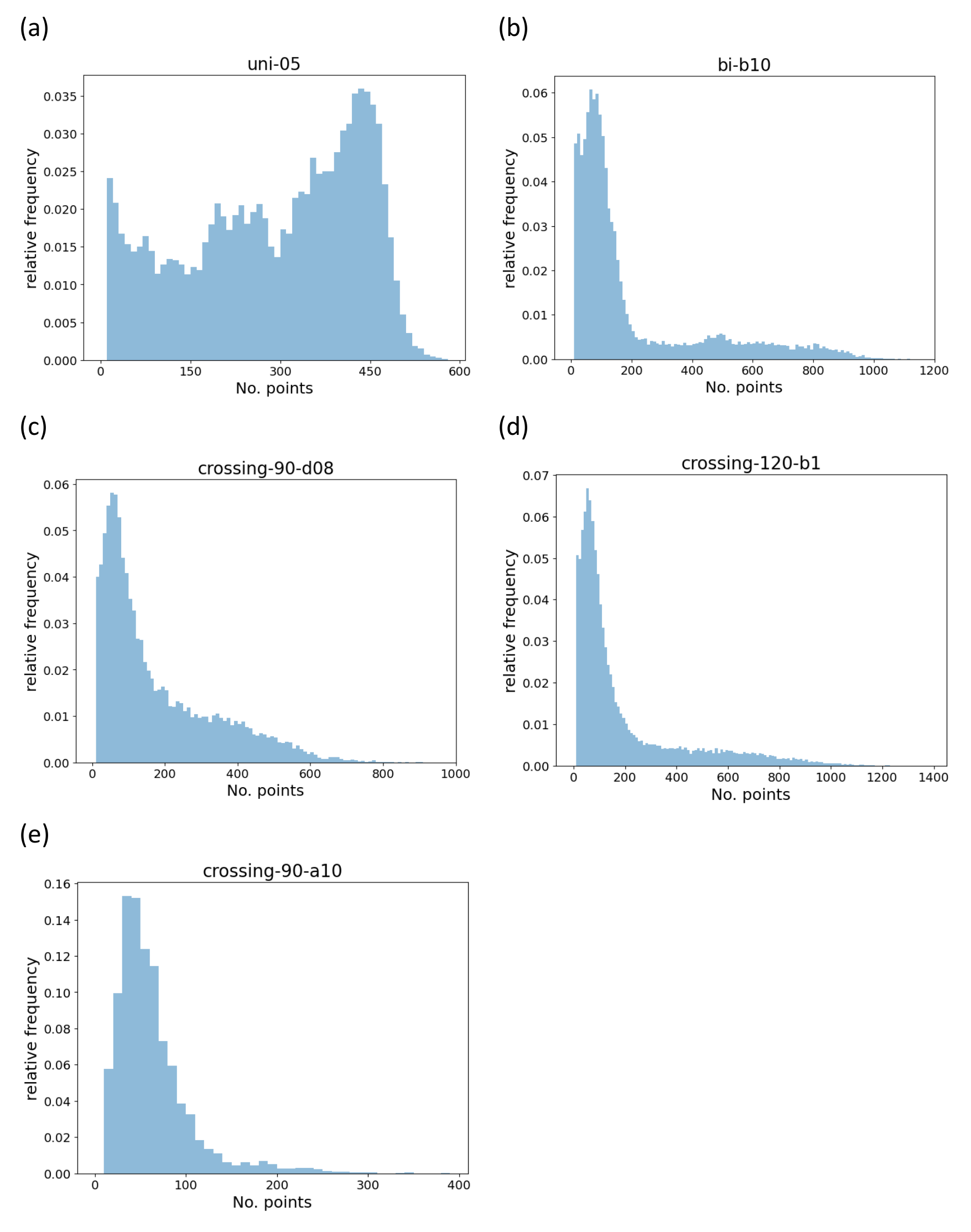}
	\caption{Histogram of the number of data points per contact interaction trajectory (a) unidirectional flow (scenario name: uni-05), (b) bidirectional flow (scenario name: bi-b01), (c) 2-way crossing flow (scenario name: crossing-90-d08), (d) 3-way crossing flow (scenario name: crossing-120-b01), and (e) 4-way crossing flow crossing flow (scenario name: crossing-90-a10).} 
	\label{fig:Hist_Npoints}
\end{figure}

From the experimental data, we extracted pairs of individuals in contact and their relative motion trajectories. We considered a pair of individuals is in contact when the two individuals are within a contact radius. The contact radius $r_c$ would depend on the form of transmission in question. In this paper, we assume a 2~m radius based on previous studies~\cite{Rutten_SciRep2022,Han_Lancet2020,Ronchi_SafetySci2020,Garcia_SafetySci2021}. For the analysis, we considered trajectories with at least 10 data points based on literature~\cite{Briane_PRE2018,Weron_PRE2019}. Table~\ref{table:BasicStatistics} shows the basic statistics of representative scenarios including the number of individuals $N$, experiment period, and the number of contacts. The number of contacts was given as the number of interacting individual pairs. Figure~\ref{fig:Hist_Npoints} presents histograms of trajectory length which is given in terms of the number of data points per contact interaction trajectory.

\section{Data Analysis}
\label{section:analysis}

Based on previous studies~\cite{Briane_2016,Briane_PRE2018,Janczura_PRE2020,Weron_PRE2019}, we evaluate a standardized value of maximal distance $T_n$ for a trajectory containing $n$ data points of position:
\begin{equation}
	T_n = \frac{D_n}{\sqrt{(t_n-t_0)\hat{\sigma}^2}},
\end{equation}
where $D_n$ is the maximal distance traveled from the initial position during the contact interaction, $t_0$ and $t_n$ are the start and end time of contact interaction, and $\hat{\sigma}$ is the consistent estimator of the standard deviation of $D_n$. The maximal distance $D_n$ is defined as
\begin{equation}
	D_n = \max_{i = 1, 2, .., n} \left\| \vec{X(t_i)}-\vec{X(t_0)} \right\|.
\end{equation}
Here, $\vec{X(t_i)}$ denotes the position at time instance $i$ and $\vec{X(t_0)}$ for the position at the start of contact interaction. The consistent estimator $\hat{\sigma}$ is given as 
\begin{equation}
	\hat{\sigma}^2 = \frac{1}{2n\Delta t} \sum_{j = 1}^{n} \left\| \vec{X(t_j)}-\vec{X(t_{j-1})} \right\|^2,
\end{equation}
where $\Delta t$ is the time step size. 

\begin{figure}[t]
	\centering
	\includegraphics[width=0.9\textwidth]{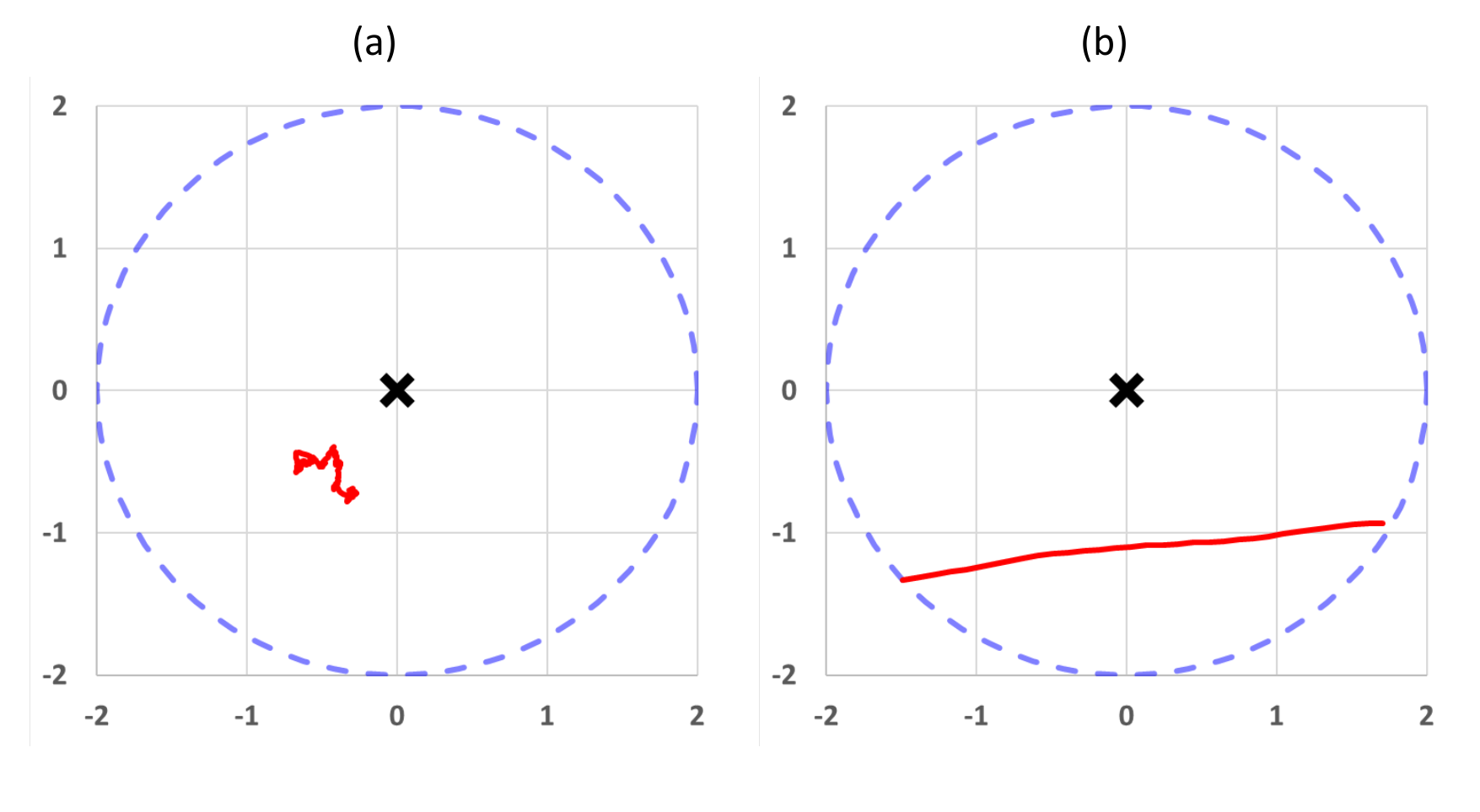}
	\vspace*{-5mm}
	\caption{Representative types of pedestrian contact interaction trajectories characterized based on the value of $T_n$. Small $T_n$ implies subdiffusive motion while large $T_n$ suggests Brownian motion: (a) Subdiffusive motion with $T_n = 0.525$ (individual id 25 and id 46 in unidirectional flow scenario uni-05) and (b) Brownian motion with $T_n = 2.5$ (individual id 65 and id 99 in 4-way crossing scenario crossing-90-a10). The position of focal individuals (id 25 in (a) and id 65 in (b)) is indicated at (0, 0) by a black cross symbol $\times$. Blue dashed circles show contact range of a focal individual ($r_c = 2$~m). The relative motion of pedestrians interacting with the focal individuals (id 46 in (a) and id 99 in (b)) denoted by red solid lines. In the lower panels, blue dashed lines indicate the ground truth trajectories of focal individuals (id 25 in (c) and id 65 in (d)), and red solid lines for pedestrians interacting with the focal individuals (id 46 in (c) and id 99 in (d)). Arrows are guide for the eyes, indicating the walking direction of individuals.} 
	\label{fig:SampleTrajectory}
\end{figure}

We can characterize different pedestrian contact interactions based on the value of $T_n$. A small $T_n$ indicates that the individuals stay close to their initial position during the contact interaction, implying subdiffusive motion in Fig.~\ref{fig:SampleTrajectory}(a). On the other hand, when individuals travel far away from their initial position during the contact interaction, one can observe large $T_n$, hinting at the possibility of Brownian motion. As can be seen from Fig~\ref{fig:SampleTrajectory}(b), individual $j$ is entering and leaving the contact circle without changing walking direction significantly. 
It should be noted that subdiffusive motion in contact interaction trajectories does not necessarily suggest that ground truth trajectories display subdiffusive motions. In the case of subdiffusive motion (see Fig.~\ref{fig:SampleTrajectory}(c)), the selected individuals move in parallel along a straight line, showing directed motions. 

We classify different motion types in line with the statistical test procedure presented in previous studies~\cite{Briane_2016,Briane_PRE2018,Weron_PRE2019}. We set Brownian motion as the null hypothesis $H_0$ and subdiffusive motion as the alternative hypothesis $H_1$. In the context of the disease spreading processes, quantifying the motion types is useful to determine whether the contact interaction is brief or long-lived, suggesting risk of virus exposure in human face-to-face interactions. In this study, subdiffusive motion is characterized by small $T_n$, implying that the risk level is high, while Brownian motion yields high $T_n$, indicating that the risk level is lower. We define a critical region based on the knowledge of $T_n$ distribution under the hypothesis $H_0$
\begin{equation}
	q_{\alpha} \leq T_n,
\end{equation}
where $q_{\alpha}$ is the quantile of $T_n$ distribution, indicating that $T_n$ lies in the critical region with the probability $1-\alpha$. We use $q_{\alpha} = 0.785$ for $\alpha= 2.5$~\% according to Refs.~\cite{Briane_2016,Briane_PRE2018,Weron_PRE2019}. 

\begin{figure}
	\centering
	\includegraphics[width=1.0\textwidth]{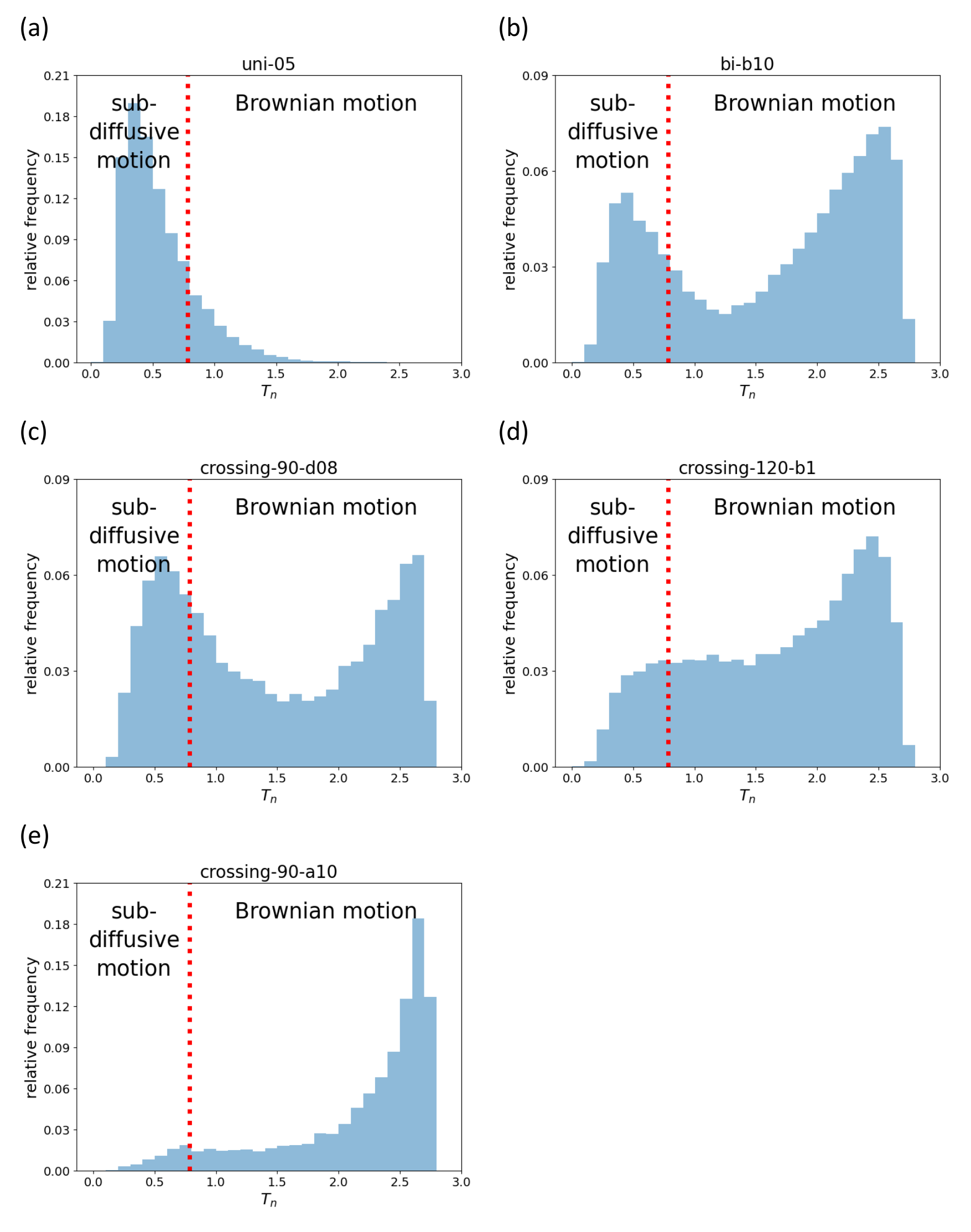}
	\caption{Histogram of the standardized maximal distance $T_n$ in trajectories (a) unidirectional flow (scenario name: uni-05), (b) bidirectional flow (scenario name: bi-b01), (c) 2-way crossing flow (scenario name: crossing-90-d08), (d) 3-way crossing flow (scenario name: crossing-120-b01), and (e) 4-way crossing flow crossing flow (scenario name: crossing-90-a10). The red dashed vertical lines indicate $\alpha$ = 2.5~\% quantile of $T_n$ distribution suggested in the previous studies~\cite{Briane_2016,Briane_PRE2018}.} 
	\label{fig:Hist_Tn0}
\end{figure}

\begin{table}
	\caption{Summary of motion type classification results.}
	\label{table:motion_type_classification}
	\setlength{\tabcolsep}{3pt}
	\centering
	\resizebox{12.5cm}{!}{
		\begin{tabular}{r*{5}{r}}
			\hline
			Setup & Scenario name & No. contacts & Subdiffusive motion & Brownian motion & \\
			\hline
			Uni-directional	& uni-05	& 25390 & 20848 (82.11\%) & 4542 (17.89\%) \\
			Bi-directional	& bi-b10	& 57126	& 14465 (25.32\%) & 42661 (74.68\%) \\
			2-way crossing	& crossing-90-d08	& 41648	& 12540 (30.11\%) & 29108 (69.89\%) \\
			3-way crossing	& crossing-120-b01	& 93156	& 14492 (15.56\%) & 78664 (84.44\%) \\
			4-way crossing	& crossing-90-a10	& 9270 	& 532 (5.74\%) & 8738 (94.26\%) \\
			\hline
		\end{tabular}
	}
\end{table}

\begin{figure}
	\centering
	\includegraphics[width=1.0\textwidth]{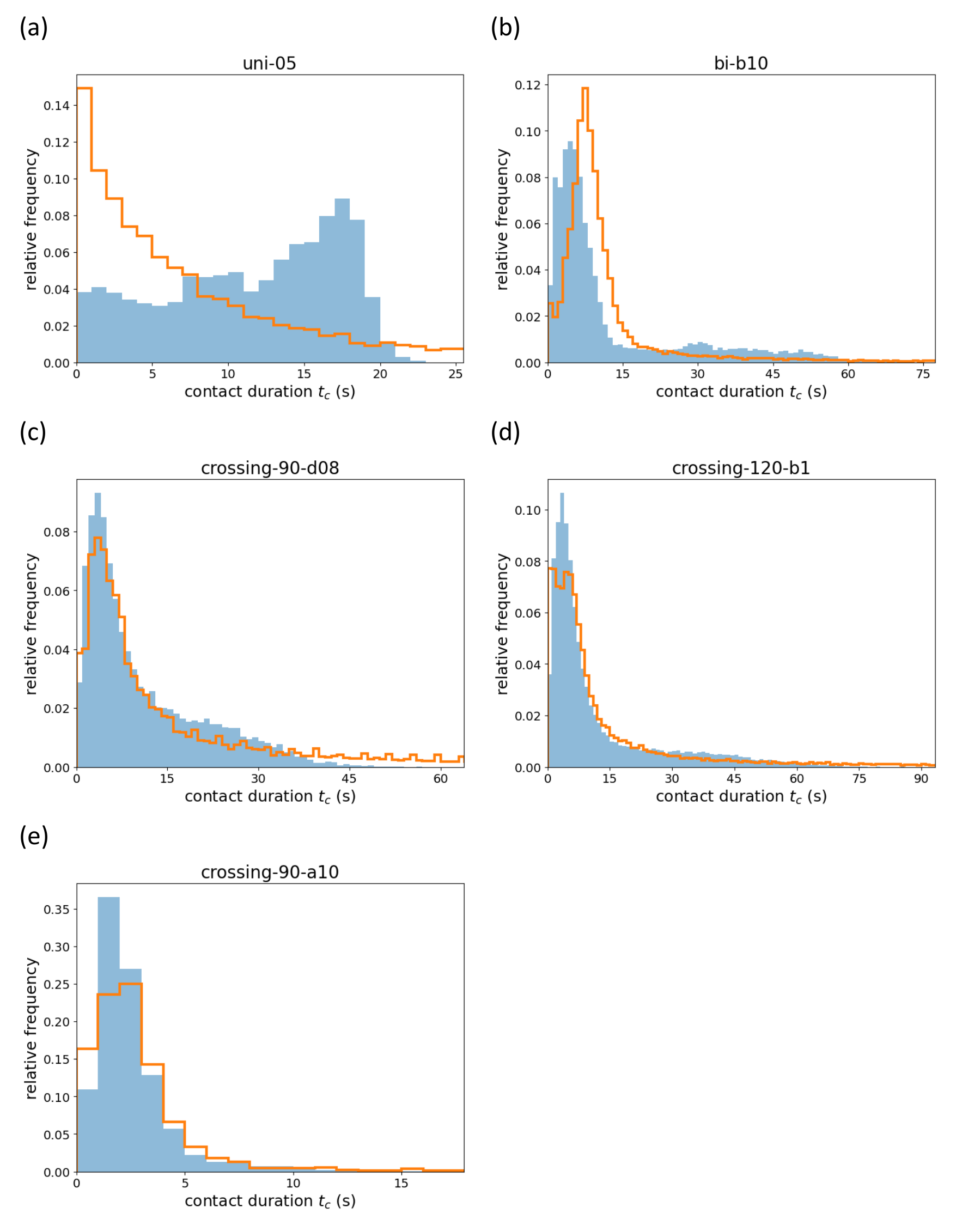}
	\caption{Histogram of the contact duration. Light blue areas indicate the histogram of actual contact duration measured from the presented experimental datasets, i.e., $t_c$. Orange lines show the histogram of the contact duration estimated based on ballistic motion assumption, i.e., $t_{c, b}$ (see Eq.~\ref{eq:t_cb}). (a) unidirectional flow (scenario name: uni-05), (b) bidirectional flow (scenario name: bi-b01), (c) 2-way crossing flow (scenario name: crossing-90-d08), (d) 3-way crossing flow (scenario name: crossing-120-b01), and (e) 4-way crossing flow crossing flow (scenario name: crossing-90-a10).} 
	\label{fig:Hist_tc}
\end{figure}

Figure~\ref{fig:Hist_Tn0} shows histograms of the standardized maximal distance $T_n$ for representative experiment scenarios. The summary of motion type classification results can be found from Table~\ref{table:motion_type_classification}. We can observe that Brownian motion is more frequently observed in most of the experiment scenarios, the exception being the unidirectional setup (scenario name uni-05). In contrast, almost all the motion in the unidirectional setup are categorized as subdiffusive motion, hinting at the possibility that the actual contact duration is much longer than one estimated based on the ballistic motion assumption. Similar to previous studies~\cite{Samar_IEEE2006,Wu_IEEE2008,Rast_PRE2022}, we estimated the contact duration of ballistic motion assumption $t_{c, b}$ as
\begin{equation} \label{eq:t_cb}
	t_{c, b} = \frac{2r_c \left| \cos \theta_i - \cos \theta_j \right|}{v_{0}(1-\cos \theta_{ij})},
\end{equation}
where $r_c = 2$~m is the contact radius and $v_{0}$ is the initial value of relative speed between individuals $i$ and $j$ measured at the beginning of contact interaction. The heading of individuals $i$ and $j$ are denoted by $\theta_i$ and $\theta_j$ respectively, and the contact angle between the individuals is given as $\theta_{ij} = \left| \theta_i - \theta_j \right|$. Similar to $v_{0}$, the heading of individuals and the contact angle between them (i.e., $\theta_i$, $\theta_j$, and $\theta_{ij}$) are measured at the beginning of contact interaction. Figure~\ref{fig:Hist_tc} illustrates the histograms of contact duration that measured from the experimental datasets $t_c$ and the estimated contact duration under the ballistic motion assumption $t_{c, b}$. In the case of the unidirectional setup, one can see a striking difference between the distributions of $t_c$ and that of $t_{c, b}$. In other setups, the distribution of $t_{c, b}$ shows good agreement with that of $t_c$.

\begin{figure}
	\centering
	\includegraphics[width=1.0\textwidth]{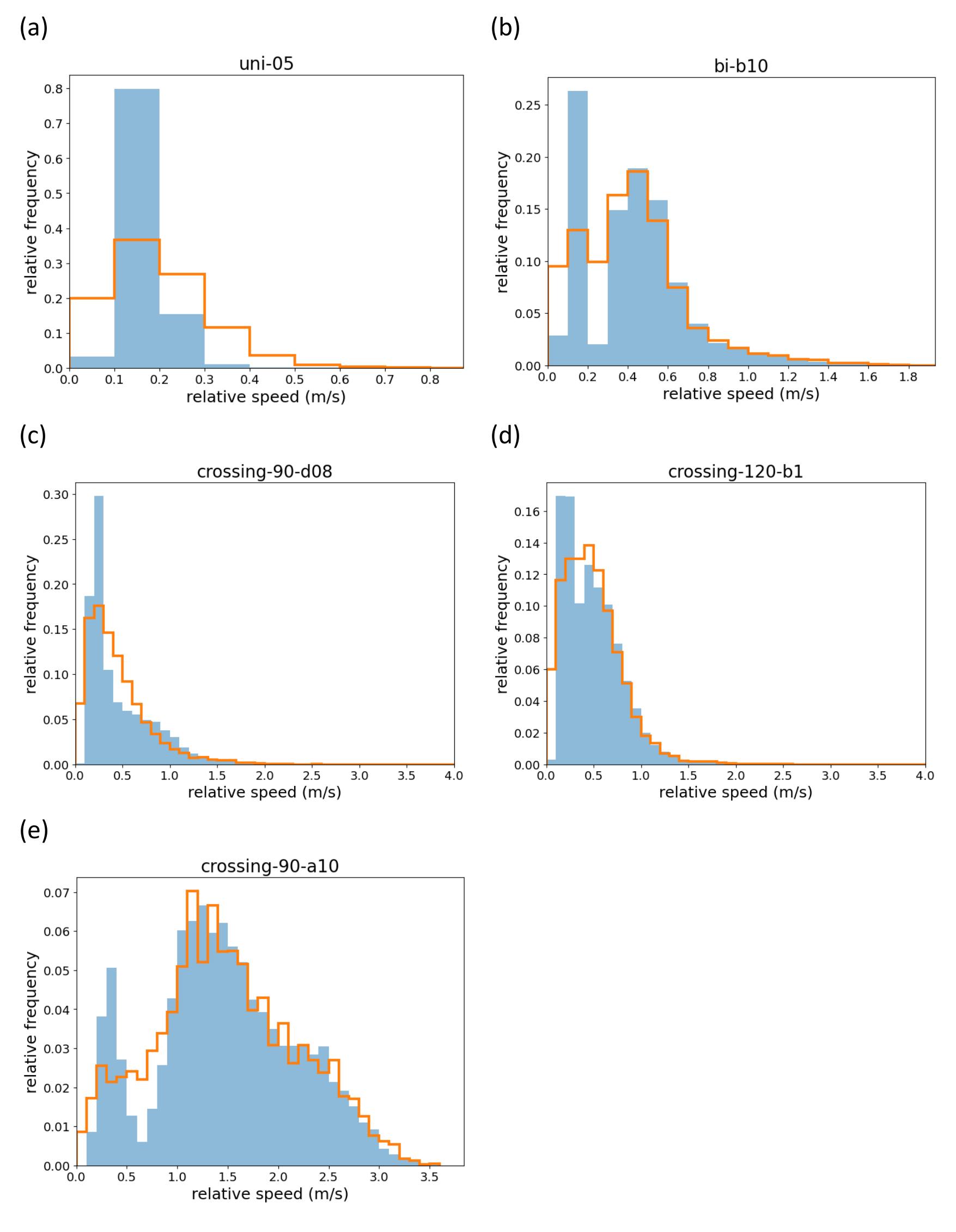}
	\caption{Histogram of the relative speed of pedestrians in contact interactions. Light blue areas indicate the histogram of actual relative speed measured from the presented experimental datasets. Orange lines are for the relative speed of contact interaction estimated based on ballistic motion assumption ($v_0$ in Eq.~\ref{eq:t_cb}). Note that the average value of relative speed during contact interaction is used for the distribution of experimental data and the initial value of relative speed $v_0$ measured at the beginning of contact interaction is used for the distribution of ballistic motion assumption. (a) unidirectional flow (scenario name: uni-05), (b) bidirectional flow (scenario name: bi-b01), (c) 2-way crossing flow (scenario name: crossing-90-d08), (d) 3-way crossing flow (scenario name: crossing-120-b01), and (e) 4-way crossing flow crossing flow (scenario name: crossing-90-a10).} 
	\label{fig:Hist_speed}
\end{figure}

To examine the reasons for the discrepancies in contact duration distributions in Fig.~\ref{fig:Hist_tc}, we compared the distributions of relative speed of pedestrians in contact interactions measured from experimental data and those generated based on the ballistic motion assumption, see Fig.~\ref{fig:Hist_speed}. Note that the average value of relative speed during contact interaction is used for the distribution of experimental data and the initial value of relative speed $v_0$ measured at the beginning of contact interaction is used for the distribution of ballistic motion assumption. As can be seen from Fig.~\ref{fig:Hist_speed}, difference in the relative frequency distribution of relative speed is significant for the case of the unidirectional flow setup, but it is less significant for other experiment setups. 

\begin{table}
	\caption{Wasserstein distance metric between relative speed distributions presented in Fig.~\ref{fig:Hist_speed}.}
	\label{table:relative_speed_Wasserstein}
	\setlength{\tabcolsep}{12pt}
	\centering
	\resizebox{12.0cm}{!}{
		\begin{tabular}{r*{4}{r}}
			\hline
			\multirow{2}{*}{Setup} & Scenario & \% Subdiffusive & Wasserstein \\
			& name & motion & distance\\			
			\hline
			Uni-directional	& uni-05	& 82.11\% & 0.0742\\
			Bi-directional	& bi-b10	& 25.32\% & 0.0145\\
			2-way crossing	& crossing-90-d08	& 30.11\% & 0.0155\\
			3-way crossing	& crossing-120-b01	& 15.56\% & 0.0065\\
			4-way crossing	& crossing-90-a10	& 5.74\% & 0.0036\\
			\hline
		\end{tabular}
	}
\end{table}

Analogous to Refs.~\cite{Chkhaidze_PLOS2019,Flam-Shepherd_NatureComm2022}, we quantify the difference in the relative frequency distributions shown in Fig.~\ref{fig:Hist_speed} by means of the Wasserstein distance. We use the 1-Wasserstein distance $W_1$ which is given as
\begin{equation}
	W_{1} = \sum_{k} \left| F(k)-F_{b}(k) \right|.
\end{equation}
Here, $F(k)$ and $F_{b}(k)$ are the cumulative distribution of relative speed relative frequency from the experimental data and the one based on the ballistic motion assumption, respectively. We compute the cumulative distribution as $F(k) = \sum_{k' \le k}^{} f(k')$, where $f(k')$ is the the relative frequency of relative speed measured for histogram bin $k'$. Table~\ref{table:relative_speed_Wasserstein} presents the Wasserstein distance measured for the presented experiment scenarios. The results show a general tendency that a higher proportion of subdiffusive motion are observed for larger values of Wasserstein distance. That is, the difference in relative speed distributions of experimental data and the ballistic motion assumption contributes considerably to the discrepancies in contact duration distributions. 

Our analysis results suggest that random walk models based on ballistic motion have limitations in accounting for the influence of pedestrian flow geometrical conditions on contact duration distributions. In the case of unidirectional flow setup, the relative speed distribution of ballistic motion assumption shows a notable difference with that of experimental data. This results in an unrealistic contact duration distribution. Furthermore, geometrical conditions yielding larger difference in the relative speed distributions tend to generate more subdiffusive motions, suggesting higher risk of disease spreading. Thus, it is desirable to have smaller difference in the relative speed distributions especially for lower relative speed.

\section{Conclusion}
\label{section:conclusion}

To examine the influence of pedestrian flow geometrical conditions on the contact duration distribution, we have analyzed pedestrian contact interaction trajectories of uni-, bi- and multi-directional flow setups in experimental data~\cite{Holl_Dissertation2016,Cao_JSTAT2017,url_dataset}. We have classified types of motions observed in the contact interactions based on standardized maximal distance $T_n$. In the unidirectional flow setup, most contact interaction trajectories have small $T_n$ values. That is, the individuals stay close to their initial position during the contact interactions, thus subdiffusive motion is frequently observed. In contrast, other experiment setups yield higher $T_n$ values. This indicates that individuals travel far away from their initial positions during the contact interactions, so Brownian motion is more frequently observed. It is noted that random walk models based on ballistic motion might not be able to generate realistic contact duration distributions depending on the geometrical conditions, especially for the case of unidirectional flow setup. This study also highlights that geometrical conditions yielding smaller difference in the relative speed distributions are preferred when diseases can be transmitted through face-to-face interactions. 

A few selected experimental scenarios have been analyzed to study the fundamental role of pedestrian flow setups (e.g., uni-, bi-, and multi-directional flow) in the distribution of pedestrian motion types and contact duration. To generalize the findings of this study, the presented analysis should be further performed with larger number of scenarios and different layouts of pedestrian facilities. Another interesting extension of the presented study can be planned in line with machine learning algorithms, for instance, developing a prediction model and performing a feature importance analysis to identify factors influencing on the motion types and duration of contact interactions~\cite{Janczura_PRE2020,Kowalek_JPhysA2022,Kowalek_PRE2019,Wagner_PLOS2017,Pinholt_PNAS2021}.

\section*{Acknowledgements}
This research is supported by Ministry of Education (MOE) Singapore under its Academic Research Fund Tier 1 Program Grant No. RG12/21 MoE Tier 1.

%
%
%
%

\end{document}